\documentclass[11pt]{article}
\begin{document}
\pagestyle{plain}
\huge
\title{\bf The two-dimensional Vavilov-\v Cerenkov radiation}
\large
\author{\bf Miroslav Pardy \\
Department of Physical Electronics \\
Masaryk University \\
Kotl\'{a}\v{r}sk\'{a} 2, 611 37 Brno, Czech Republic\\
email:pamir@physics.muni.cz}
\date{\today}
\maketitle
\vspace{50mm}
\large
\begin{abstract}
\large

We derive the power spectrum of photons generated by charged particle moving in parallel direction to the graphene-like structure with index of refraction n. Some graphene-like structures, for instance graphene with implanted ions, or, also 2D-glasses, are dielectric media, and it means that it enables the experimental realization of the Vavilov-\v Cerenkov radiation. We calculate it from the viewpoint of the Schwinger theory of sources.
\end{abstract}
\vspace{7mm}

\baselineskip 15pt

The fast moving charged particle
in a medium when its speed is faster than the speed of light in this medium produces electromagnetic radiation which is called the Vavilov-\v{C}erenkov radiation. 

The prediction of Cerenkov radiation came long ago. Heaviside (1889)
investigated the possibility of a charged object moving
in a medium faster than electromagnetic waves in the same
medium becomes a source of directed electromagnetic radiation. Kelvin (1901) presented an idea that the emission of particles is possible at a speed greater than that of light. Somewhat later, Sommerfeld (1904) proposed the hypothetical radiation with a sharp angular distribution. However, in fact, from experimental point of view, the electromagnetic \v Cerenkov
radiation was first observed in the early 1900's by experiments developed by Marie and Pierre Curie when studying radioactivity emission.  In essence they observed the  emission of a bluish-white light from transparent substances in the neighborhood of strong radioactive source. But the
first attempt to understand the origin of this was made by Mallet (1926, 1929a, 1929b) who observed that the light emitted by a variety of transparent bodies placed close to a radioactive source always had the same bluish-white quality, and
that the spectrum was continuous, with no line or band structure
characteristic of fluorescence. 

Unfortunately,
these investigations were forgotten for many years.
\v Cerenkov experiments (\v{C}erenkov, 1934) was performed at the suggestion of Vavilov who opened a door to the true physical nature of the this effect\footnote{So, the adequate name of this effect is the Vavilov-\v Cerenkov effect. In the English literature, however, it is usually called the \v Cerenkov effect.} (Bolotovskii, 2009).

This radiation was first  theoretically interpreted by Tamm  and Frank (1937) in the framework of the classical electrodynamics. The source theoretical description of this effect was given by Schwinger et~al. (1976) at the zero temperature regime and the classical spectral formula was generalized to the  finite temperature situation and for the massive photons by autor (Pardy, 1989; 2002). The Vavilov-\v Cerenkov effect was also used by author (Pardy, 1997) to possible measurement of the Lorentz contraction.
 
Let us start with the  three dimensional source theory  formulation  of the problem. Source theory (Schwinger et al., 1976) is the theoretical construction which uses
quantum-mechanical particle language. Initially it was constructed
for description of the particle physics situations occurring in the
high-energy physics experiments. However, it was found that the
original formulation simplifies the calculations in the
electrodynamics and gravity where the interactions are mediated by
photon or graviton respectively.

The basic formula in the source theory is the vacuum to vacuum
amplitude:

$$<0_{+}|0_{-}> = e^{\frac {i}{\hbar}\*W(S)},\eqno(1)$$
where the minus and plus tags on the vacuum symbol are causal
labels, referring to any time before and after space-time region
where sources are manipulated. The exponential form is introduced
with regard to the existence of the physically independent
experimental arrangements which has a simple consequence that the
associated probability amplitudes multiply and corresponding
$W$ expressions add.

The electromagnetic field is described by the amplitude (1) with the action

$$W(J) = \frac {1}{2c^2}\*\int (dx)(dx')J^{\mu}(x)
\*D_{+\mu\nu}(x-x')J^{\nu}(x'),\eqno(2)$$
where the dimensionality of $W(J)$ is the same as the dimensionality
of the Planck constant $\hbar$. $J_{\mu}$ is the charge and current densities, where quantity  $J_{\mu}$ is conserved.
The symbol $D_{+\mu\nu}(x-x')$, is the photon
propagator and its explicit form will be determined later.

It may be easy to show that the probability of the persistence
of vacuum is given by the following formula (Schwinger et al., 1976):

$$|<0_{+}|0_{-}>|^2 = \exp\{-\frac {2}{\hbar}{\rm Im}\, W\} \,
\stackrel{d}{=}\, \exp\{-\int\,dtd\omega
\frac {P(\omega,t)}{\hbar\omega}\},\eqno(3)$$
where we have introduced the so called power spectral
function $P(\omega,t)$ (Schwinger et al., 1976). In order to extract this
spectral function from ${\rm Im}\, W$, it
is necessary to know the explicit form of the photon propagator
$D_{+\mu\nu}(x-x')$.

The electromagnetic field is
described by the four-potentials $A^{\mu}(\varphi,{\bf A })$ and it is generated, including a particular choice of gauge, by the four-current $J^{\mu}(c\varrho,{\bf J})$ according to the differential equation,  (Schwinger et al., 1976):

$$(\Delta-\frac{\mu\varepsilon}{c^2}\*\frac{\partial^2}{\partial\*t^2})A^{\mu}
= \frac{\mu}{c}(g^{\mu\nu}+\frac
{n^2-1}{n^2}\eta^\mu\*\eta^\nu)J_{\nu}
\eqno(4)$$
with the corresponding Green function $D_{+\mu\nu}$:

$$D_{+}^{\mu\nu} = \frac{\mu}{c}\*(g^{\mu\nu}+\frac{n^2-1}{n^2}\*
\eta^{\mu}\eta^{\nu})\*D_{+}(x-x'),\eqno(5)$$
where $\eta^{\mu}\equiv (1,{\bf 0})$,
$\mu$ is the magnetic permeability of the dielectric medium
with the dielectric constant $\varepsilon$, $c$ is the velocity of
light in vacuum, $n$ is the index of refraction of this medium, and
$D_{+}(x-x')$ was derived by Schwinger et al. (1976) in the following form:

$$D_{+}(x-x') =\frac {i}{4\pi^2\*c}\*\int_{0}^{\infty}d\omega
\frac {\sin\frac{n\omega}{c}|{\bf x}-{\bf x}'|}{|{\bf x} - {\bf x}'|}\*
e^{-i\omega|t-t'|}.\eqno(6)$$

Using formulas (2), (3), (5) and (6), we get
for the power spectral formula the following expression (Schwinger et al., 1976):

$$P(\omega,t) = -\frac {\omega}{4\pi{^2}}\frac {\mu}{n^{2}}
\int\, d{\bf x}d{\bf x}'dt'\*
\frac {\sin\frac{n\omega}{c}|{\bf x}-{\bf x}'|}{|{\bf x}-{\bf x}'|}\*
\cos[\omega(t- t')]  \times $$

$$\times\;\left\{\varrho({\bf x},t)\varrho({\bf x'},t') - \frac{n^{2}}{c^2}\*
{\bf J}({\bf x},t)\cdot{\bf J}({\bf x'},t')\right\}.\eqno(7)$$

Now, we are prepared to apply the last formula to the
situations of the two dimensional dielectric medium. We derive here the power spectrum of photons generated by charged particle moving in parallel direction to the graphene-like structure with index of refraction n. While the graphene sheet is conductive, some graphene-like structures, for instance graphene with implanted ions, or, also 2D-glasses, are dielectric media, and it means
that it enables the experimental realization of the Vavilov-\v Cerenkov radiation. Some graphene-like structure can be represented by graphene-based polaritonic crystal sheet (Bludov et al., 2012) which can be used to study the Vavilov-\v Cerenkov effect. We calculate it from the viewpoint of the Schwinger theory of sources.

 The charge and current density of electron
moving with the velocity ${\bf v}$ and charge $e$ is as it is well known:

$$\varrho = e\delta({\bf x}-{\bf v}t)\eqno(8)$$

$${\bf J} = e{\bf v}\delta({\bf x}-{\bf v}t)\eqno(9)$$

In case of the the two dimensional Vavilov-\v Cerenkov radiation  by source theory  formulation, the form of equations (2) and (3) is the same with the difference that $\eta^{\mu}\equiv(1,{\bf 0})$ has two space components, or $\eta^{\mu}\equiv(1,0,0)$, and the Green function $D_{+}$ as the propagator must be determined by the two dimensional procedure. In other words, the Fourier form of this propagator is with $(dk) = dk^{0}d{\bf k} = dk^{0}dk^{1}dk^{2} = dk^{0}kdkd\theta$

$$D_{+}(x-x') =\int\frac {(dk)}{(2\pi)^3}\,\frac{1}{{\bf k}^{2} -n^{2}(k^{})^{2}}\, e^{ik(x - x')},\eqno(10)$$
or, with $R = |{\bf x} - {\bf x}'|$ 

$$D_{+}(x-x') =\frac {1}{(2\pi)^{3}}\int_{0}^{2\pi}d\theta\int_{0}^{\infty}kdk\int_{-\infty}^{\infty}\frac{d\omega}{c}
\frac {e^{ikR\cos\theta -i\omega(t-t')}}
{k^{2} - \frac{n^{2}\omega^{2}}{c^{2}} -i\varepsilon}.\eqno(11)$$

Using $\exp(ikR\cos\theta) = \cos(kR\cos\theta) + i\sin(kR\cos\theta)$ and ($z = kR$) 

$$\cos(z\cos\theta) = J_{0}(z) + 2\sum_{n=1}^{\infty}(-1)^{n}J_{2n}(z)\cos2n\theta \eqno(12)$$
and 

$$\sin(z\cos\theta)= \sum_{n=1}^{\infty}(-1)^{n}J_{2n-1}(z)\cos(2n-1)\theta ,\eqno(13)$$
where $J_{n}(z)$ are the Bessel functions (Kuznetsov, 1962), we get after integration over $\theta$:

$$D_{+}(x-x') =\frac {1}{(2\pi)^2}\int_{0}^{\infty}kdk\int_{-\infty}^{\infty}\frac{d\omega}{c}
\frac {J_{0}(kR)} {k^{2} - \frac{n^{2}\omega^{2}}{c^{2}} -i\varepsilon}e^{-i\omega(t-t')}\eqno(14)$$
and it is pedagogically useful to say that the Bessel function $J_{0}(z)$ has the following expansion (Kuznetsov, 1962):

$$J_{0}(z)= \sum_{s = 0}^{\infty}\frac{(-1)^{s}z^{2s}}{s!s!2^{2s}},\eqno(15)$$
which is convergent for all $z$ with regard to the d'Alembert convergence criterion.

The $\omega$-integral in (14) can be performed using the residuum theorem after integration in the complex half $\omega$-plane.

The result of such integration is the propagator $D_{+}$ in the following form:

$$D_{+}(x-x') =\frac {i}{2\pi c}\int_{0}^{\infty}d\omega
J_{0}\left(\frac{n\omega}{c}|{\bf x} - {\bf x}'|\right)e^{-i\omega|t-t'|}.\eqno(16)$$

The initial terms of the expansion of the Bessel function with  zero index is as follows:

$$J_{0}(z) = 1 - \frac{z^{2}}{2^{2}} + \frac{z^{4}}{2^{2}4^{2}} - \frac{z^{6}}{2^{2}4^{2}6^{2}} + \frac{z^{8}}{2^{2}4^{2}6^{2}8^{2}} - ...\eqno(17)$$

The spectral formula for the two dimensional Vavilov-\v Cerenkov radiation is the analogue of the formula (7), or,

$$P(\omega, t) = -\frac {\omega}{2\pi}\frac {\mu}{n^{2}}
\int\, d{\bf x}d{\bf x}'dt'\*J_{0}\left(\frac{n\omega}{c}|{\bf x} - {\bf x}'|\right)\*
\cos[\omega(t- t')] \times $$

$$\times\;\left\{\varrho({\bf x},t)\varrho({\bf x'},t') - \frac{n^{2}}{c^2}\*
{\bf J}({\bf x},t)\cdot{\bf J}({\bf x'},t')\right\}\eqno(18)$$
where the charge density and current involves only two dimensional velocities and integration is also only two dimensional. 

The difference is in the replacing mathematical formulas as follows:

$$\frac{\sin\frac{n\omega}{c}|{\bf x}-{\bf x}'|}{|{\bf x}-{\bf x}'|}
\quad \longrightarrow \quad J_{0}\left(\frac{n\omega}{c}|{\bf x} - {\bf x}'|\right)
\eqno(19)$$

So, After insertion the quantities (8) and (9) into (18), we get:

$$P(\omega,t) = \frac {e^{2}}{2\pi}\frac {\mu\omega v}{c^{2}}\left(1 - \frac{1}{n^{2}\beta^{2}}\right)
\int\, dt'\*
J_{0}\left(\frac{nv\omega}{c}|t-t'|\right)
\cos[\omega(t- t')], \quad \beta = v/c,\eqno(20)$$
where the $t'$-integration must be performed. Putting $\tau = t'-t$,
we get the final formula:

$$P(\omega,t) = \frac {e^{2}}{2\pi}\frac {\mu\omega v}{c^{2}}\left(1 - \frac{1}{n^{2}\beta^{2}}\right)
\int_{-\infty}^{\infty}\, d\tau\*
J_{0}\left(n\beta\omega\tau\right)
\cos(\omega\tau), \quad \beta = v/c.\eqno(21)$$

The integral in formula (21) is involved in the tables of integrals (Gradshteyn et al. 1962) on page 745, no. 8. Or,

$$J = \int_{0}^{\infty}\, dx\* J_{0}\left(ax\right)
\cos(bx) = \frac{1}{\sqrt{a^{2}-b^{2}}}, \quad 0<b<a $$

$$ J = \infty, a= b; \; J = 0,\quad  0<a<b \eqno(22)$$

In our case we have $a = n\beta\omega$  and $b = \omega$. So, the power spectrum in eq. (21) is as follows with $J_{0}(-z) = J_{0}(z)$:

$$P = \frac {e^{2}}{\pi}\frac {\mu v}{c^{2}}\left(1 - \frac{1}{n^{2}\beta^{2}}\right)\* \frac{2}{\sqrt{n^{2}\beta^{2}- 1}},\quad n\beta > 1, \; \beta = v/c.\eqno(23)$$
and 

$$P = 0; \; n\beta < 1 ,\eqno(24)$$
which means that the physical meaning of the quantity $P$ is really the Vavilov-\v Cerenkov radiation. And it is in our case the two dimensional form of this radiation.
 
While the formula for the three dimensional (3D) Vavilov-\v Cerenkov radiation  is well known from textbooks and monographs, the two dimensional  (2D) form of the Vavilov-\v Cerenkov radiation was derived here. Let us remember, in conclusion, the fundamental features of the 3D Vavilov-\v Cerenkov radiation:\\
1) The radiation arises only for particle velocity greater than the velocity of light in the dielectric medium.\\
2) It depends only on the charge and not on mass of the moving particles\\
3) The radiation is produced in the visible interval of the light frequencies and partly in the ultraviolet part of the frequency spectrum. The radiation does not exists for very short waves because from the theory of index of refraction $n$ it follows that $n<1$ in a such situation.\\
4) The spectral dependency on the frequency is linear for the 3D homogeneous medium. \\
5) The radiation generated in the given point of the trajectory spreads on the surface of cone with the vertex in this point and with the axis identical with the direction of motion of the particle. The vertex angle of the cone is given by the relation $\cos\Theta = c/nv$. There is no cone in the 2D dielectric medium

Let us remark that the energy loss of a particle caused by the Vavilov-\v Cerenkov radiation are approximately equal to 1\% of all energy losses in the condensed
matter such as the bremsstrahlung and so on. The fundamental importance of the Vavilov-\v Cerenkov radiation is in its use for the modern detectors of very speed charged particles in the high energy physics. The detection of the Vavilov-\v Cerenkov radiation enables to detect not only the existence of the particle, however, also the direction of motion and its velocity and according also its charge. The two-dimensional Vavilov-\v Cerenkov radiation was still not applied, nevertheless it is promising.

\vspace{7mm}
{\bf REFERENCES}
\vspace{5mm}
\begin{flushleft}

Bludov, Yu. V., Peres, N. M. R. and Vasilevskiy, M. I. (2012). Graphene-based polaritonic crystal, arXiv:1204.3900v1,[cond-mat.mes-hall]. \\
Bolotovskii, B. M. (2009). Vavilov-Cherenkov radiation: its discovery and application, Physics - Uspekhi {\bf 52}(11) 1099 - 1110.\\
\v Cerenkov, P. A. (1934). The visible radiation of pure liquids caused by $\gamma$-rays, {\it Comptes Rendus Hebdomaclaires des Scances de l' Academic des Sciences USSR} {\bf 2}, 451.\\
Gradshteyn, J. S. and  Ryzhik, I. M. (1962). {\it Tables of integrals, sums, series and products}, Moscow. (in Russian).\\
Heaviside, O. (1889). On the electromagnetic effects due to the motion of electrification through a dielectric, Philos. Mag., S. 5, {\bf 27}, 324– 339. \\
Kelvin, L. (1901). Nineteenth century clouds over the dynamical theory of heat and light, Philos. Mag., S. 6, {\bf 2}, 1–40. \\
Kuznetsov, D. S. {\it The special functions}, Moscow. (in Russian).\\
Mallet, L.  (1926). Spectral research of luminescence of water and other media with gamma radiation, Comptes Rendus, {\bf 183}, 274.;  ibid. (1929a). Comptes Rendus, {\bf 187}, 222.; ibid. (1929b). Comptes Rendus, {\bf 188}, 445.\\
Pardy, M. (1989). Finite-temperature \v Cerenkov radiation, {\it Physics Letters A} {\bf 134}(6), 357.\\
Pardy, M. (1997). \v Cerenkov effect and the Lorentz contraction, {\it Phys. Rev. A} {\bf 55}, 1647.\\
Pardy, M. (2002). \v Cerenkov effect with massive photons, {\it International Journal of Theoretical Physics}, {\bf 41}(5), 887.\\
Schwinger, J., Tsai, W. Y. and Erber, T. (1976). Classical and quantum theory of synergic synchrotron-\v Cerenkov radiation, {\it Annals of Physics (NY)} {\bf 96}, 303.\\
Sommerfeld, A. (1904). Zur Elektronentheorie: II. Grundlagen f{\"u}r eine allgemeine Dynamik des Elektrons, G{\"o}ttingen Nachr., {\bf 99}, 363-439.\\
Tamm, I. E. and Frank, I. M. (1937). The coherent radiation of a fast electron in a medium, {\it Dokl. Akad. Nauk SSSR} {\bf 14}, 109.
\end{flushleft}
\end{document}